\documentclass[
superscriptaddress,
preprint,
reprint,
showpacs,preprintnumbers,
 amsmath,amssymb,
 aps,
prl,
]{revtex4-1}
\usepackage{appendix}
\usepackage{SIunits}
\usepackage{sistyle}
\usepackage[utf8x]{inputenc}
\usepackage{graphicx}   
\graphicspath{{fig/}}   
\usepackage{dcolumn}
\usepackage{bm}
\usepackage{hyperref}
\usepackage{epstopdf}
\usepackage{lipsum}
\usepackage{braket}
\DeclareGraphicsExtensions{.pdf,.png,.jpg,.jpeg,.mps}
\usepackage{pgf}
\usepackage{tikz}

\usepackage{changes}
\newcommand{\figref}[1]{Fig.~\ref{#1}}
\newcommand{\tabref}[1]{Tab.~\ref{#1}}

\newcommand{\eqnref}[1]{Eq.~\eqref{#1}}

\newcommand{\akuts}[1]{``#1''}

\newcommand{\sous}[1]{\ensuremath{_{\textrm{#1}}}}

\begin{document}


\title{Control of the Coupling Strength and Linewidth of a Cavity-Magnon Polariton  }

\author{Isabella Boventer}
\affiliation{Institute of Physics, Johannes Gutenberg University Mainz, 55099 Mainz, Germany}
\affiliation{Institute of Physics, Karlsruhe Institute of Technology, 76131 Karlsruhe, Germany}
\author{Christine Dörflinger}
\affiliation{Institute of Physics, Karlsruhe Institute of Technology, 76131 Karlsruhe, Germany}
\author{Tim Wolz}
\affiliation{Institute of Physics, Karlsruhe Institute of Technology, 76131 Karlsruhe, Germany}
\author{Rair Mac{\^e}do}
\affiliation{School of Engineering, Electronics \& Nanoscale Engineering Division, University of Glasgow, Glasgow G12 8QQ, United Kingdom}

\author{Romain Lebrun}
\affiliation{Institute of Physics, Johannes Gutenberg University Mainz, 55099 Mainz, Germany}
\author{Mathias Kl\"aui}
\email{Klaeui@Uni-Mainz.de}
\affiliation{Institute of Physics, Johannes Gutenberg University Mainz, 55128 Mainz, Germany}
\affiliation{Materials Science in Mainz, Johannes Gutenberg University Mainz, 55128 Mainz, Germany}
\author{Martin Weides}
\affiliation{Institute of Physics, Karlsruhe Institute of Technology, 76131 Karlsruhe, Germany}
\affiliation{School of Engineering, Electronics \& Nanoscale Engineering Division, University of Glasgow, Glasgow G12 8QQ, United Kingdom}


\date{\today}

\begin{abstract}
The full coherent control of hybridized systems such
as strongly coupled cavity photon-magnon states is a crucial step to enable future information processing technologies. Thus, it is particularly interesting to engineer deliberate control mechanisms such as the full control of the coupling strength as a measure for coherent information exchange. 
In this work, we employ cavity resonator spectroscopy to demonstrate the complete control of the coupling strength of hybridized cavity photon-magnon states. For this, we use two driving microwave inputs which can be tuned at will. 
Here, only the first input couples directly to the cavity resonator photons, whilst the second tone exclusively acts as a direct input for the magnons.   
For these inputs, both the relative phase $\phi$ and amplitude $\delta_0$ can be independently controlled.   
 We demonstrate that for specific quadratures between both tones we can increase the coupling strength, close the anticrossing gap, and enter a regime of level merging.   
At the transition, the total amplitude is enhanced by a factor of 1000 and we observe an additional linewidth decrease of $13\%$ at resonance due to level merging.   
Such control of the coupling, and hence linewidth, open up an avenue to enable or suppress an exchange of information and bridging the gap between quantum information and spintronics applications.
\end{abstract}


\maketitle
Polaritons are the quasiparticles associated with the coupling of electromagnetic waves with an excited state of matter \cite{Mills1974, Litinskaya2016}.
Such hybridized systems are promising candidates
for applications as they can combine the advantages of the different physical systems and overcome the limitations of a single one \cite{Kurizki2015, Kubo_2011, Morton2011}. 
While hybrid quantum circuits represent a tool for the deliberate control of quantum states,  light-matter interactions can be thought as an equivalent for macroscopic systems through various types of polaritons such as exciton-photon, or magnon polaritons (MPs) \cite{Blatt2008, Hanson2008, Georgescu2014, Aspelmeyer2014, Angelakis2017, Yamamotoa1994, Pitarke2006, Barnes2003}. For instance, MPs enable examining the spin-photon interaction, where the magnons are the associated quanta of a collective spin excitation \cite{MelkovBook}. Thus, the study and manipulation of spin-photon interaction could lead to the development of spintronic applications \cite{Zutic2004, Tabuchi2015, Lu2016, Baltz2018, Chumak2017, Harder2018}. 
However, the realization of applications based on such hybrid systems also requires full control over the coupling strength $g_{\mathrm{eff}}$ which  
is a measure for the coherent exchange of information. Therefore, such control would enable a deliberate enhancement or suppression of the information exchange \cite{9783540285731}. This is of broad interest, and has been studied for various systems such as single atoms, optomechanical circuits, exciton or surface plasmon polaritons, and quantum dots strongly coupled to a nanocavity \cite{Choi2010, Bernier2018, Juggins2018, Moilanen2017, Dory2016}.\newline\\
In the field of cavity photon-magnon polariton (CMP) spectroscopy, various recent experiments studied MPs using a Yttrium-Iron-Garnet (YIG) sphere as the magnonic sample in a cavity resonator driven with \textit{a single} microwave tone. 
At resonance, the photon states fully hybridize with the magnon states, and a CMP is created \cite{Bai2015}. In the strong coupling regime, the coupling strength is related to the anticrossing gap by $\Delta \omega=2g_{\mathrm{eff}}$ and depends on the resonator geometry and the selected sample \cite{Tabuchi_2014, Zhang_2014}.
CMPs have been well studied for different configurations and temperatures throughout the last years \cite{Tabuchi2015, Bai2015, Harder_2016, Haidar_2015, Cao2015, Yao2017, Zhang2015, Zhang2017a, Harder2018, Yang2019, Liensberger2019}. 
For instance, long-distance manipulation of spin current has been demonstrated using strong cavity photon-magnon coupling in cavity resonators \cite{Bai2017}. 
At room temperature, the origin of the coherent cavity photon-magnon coupling can be attributed to a fixed phase correlation of the electromagnetic fields as shown by Bai et al. \cite{Bai2015}. 
Consequently, by finding a possibility to tune the phase relation between the cavity photon and the magnon, $g_{\mathrm{eff}}$ could be manipulated. This has recently shown to be possible when changing the position of the sample inside the cavity \cite{Harder2018a}. However, the introduction of a second field, driving only the magnons, could also achieve similar behavior as theorized by Grigoryan et al. \cite{Grigoryan2018}.
\\\\
Here, we report on the experimental realization of a tuning mechanism for $g_{\mathrm{eff}}$ for a three dimensional (3D) system.  
We show both the maximization of the coupling and complete merging of the cavity photon and magnon states. 
This is realized by the introduction of a second tone which only drives the magnons and controlling its relative phase $\phi$ and amplitude ratio $\delta_0$ with respect to the cavity field from the first tone. 
The anticrossing gap at resonance can be set to zero by simply controlling the inputs. This yields a new regime for the CMP which we call \akuts{level merging}. 
We also study the impact of changing $\phi$ and $\delta_0$ on the linewidth and total amplitude. 
In contrast to other works, our approach is entirely externally tunable. It is not necessary to modify the experimental environment such as changing the position of the magnonic sample in the cavity resonator \cite{Harder2018a, Yao2019}. In addition, we show this external control for a 3D hybrid cavity photon-magnon system, not 2D \cite{Bhoi2019}, which could be adapted to similar 3D hybrid systems. 
 \\\\
\begin{figure}[t!]
 \centering
 \includegraphics[scale=0.3]{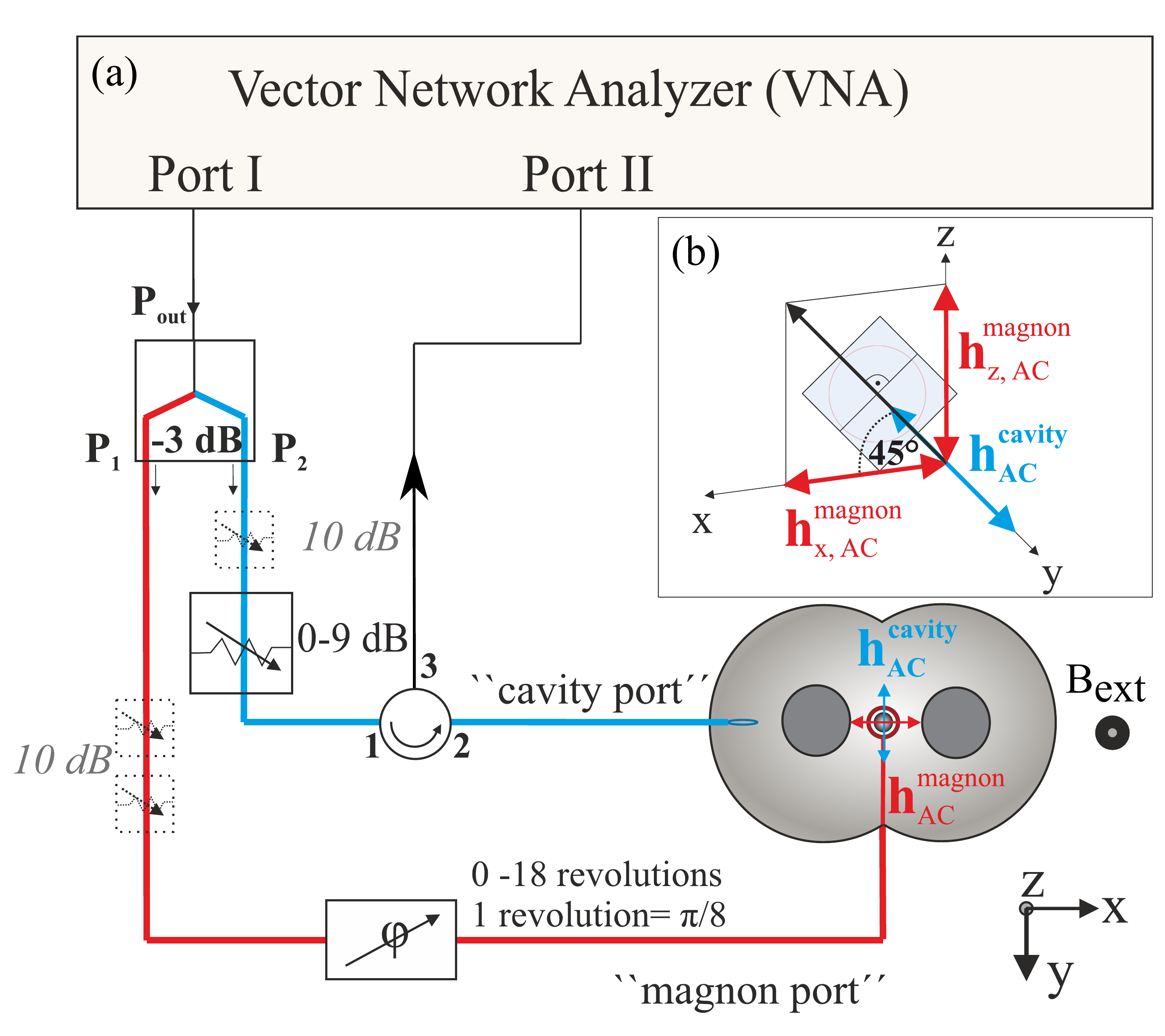}
 \caption{(a) Schematics of the experimental set up showing the coherent signal from Port I [output power level  $P_{\mathrm{out}}=-5\,\mathrm{dBm} \, (0.3\,\mathrm{mW})$] being divided by a power splitter.
The value for $\delta_{\mathrm{0}}$ is controlled by a variable ($0\,\mathrm{to}\,9\,\mathrm{dB}$) permanently inserted attenuator in the path of the cavity port and, if necessary, fixed attenuators ($10\,\mathrm{dB}$ each).  
A mechanically tunable phase shifter in the path of the magnon port modulates the phase (systematic uncertainty of  $\pm 0.02\pi/8$). 
The system's response is measured in reflection at Port II of the VNA. 
(b) Orientation of the coupling loop tilted by $45^\circ$ to the $xy$-plane around the YIG sphere and relative orientation of the AC magnetic fields from the cavity and the magnon port. }
 \label{Fig.1:Setup}
 \end{figure}
Our experiment consists of a commercially bought YIG sphere ($\rm{Y_{3}Fe_{5}O_{12}}, r=0.1\,\mathrm{mm}$) \cite{ferrisphere}, placed in the antinode of the \akuts{brights} mode magnetic AC field of a reentrant cavity resonator with $\omega_c/2\pi =6.50\,\mathrm{GHz}$ \cite{ZareRameshti2015,Goryachev_2014,Boventer2018, Pfirrmann2019}. 
Here, the additional input, called magnon port, is comprised of a metallic loop around the YIG sphere.  
Our observations are due to the superposition of two tones out of phase with one another, as predicted by Grigoryan et al.  \cite{Grigoryan2018}. However, as the loop is tilted by $\chi=45^\circ$ with respect to the cavity's $xy$-plane, we have used a different geometry based on a tilted coupler topology. It was experimentally found that the crosstalk between the ports is hereby minimized, which does not prevent from a qualitative agreement as we later demonstrate. 
After an initial adjustment, the orientations of the loops of both ports remain fixed. In order to obtain two coherent microwave drives up to this phase, we use the first port of a vector network analyzer (VNA) with an output power level at the VNA of $P_{\mathrm{out}}=-5\,\mathrm{dBm} \, (0.3\,\mathrm{mW})$ as the only microwave source in the system. 
The signal is then split into two paths with a mechanically tunable phase shifter added to the path of the magnon port. The
relative amplitude $\delta_0$ is varied by the variable and fixed attenuators in path $P_1$ and $P_2$, respectively [c.f. \figref{Fig.1:Setup}(a)]. 
The data itself is taken in reflection  [$S_{11}(\omega)$] from the cavity port and recorded at Port II of the VNA. 
\\\\
Since magnons are quasiparticles associated with a collective spin excitation in a magnetic material \cite{Kittel_1948, ZareRameshti2015}, the time-varying magnetic fields, inside our cavity resonator, drive the spins in the YIG sphere out of their equilibrium orientation. Their time evolution can be described by the Landau-Lifshitz-Gilbert [LLG] equation \cite{Morrish2001}:
$\frac{\partial \textbf{\emph{M}}}{\partial t}=\gamma \textbf{\emph{M}} \times \textbf{\emph{H}}_{\rm{eff}} -\frac{\alpha}{M_s} \left(\mathbf{M}\times \frac{\partial \textbf{\emph{M}}}{\partial t}\right)$
where $\textbf{\emph{H}}_{\rm{eff}}$ denotes the effective magnetic field experienced by the spins, $\gamma$ the gyromagnetic ratio, $\alpha$ the Gilbert damping parameter and $\mathrm{M}_s$ the saturation magnetization. 
In our particular case, we can quote the effective field as $\textbf{\emph{H}}_{\rm{eff}}=\textbf{\emph{H}}_{\rm{ext}}+\textbf{\emph{h}}_{\mathrm{AC}}^{\mathrm{cavity}}+\textbf{\emph{h}}_{\mathrm{AC}}^{\mathrm{magnon}}$, where $\textbf{\emph{H}}_{\rm{ext}}=(0,0,H_{\mathrm{ext}})$ describes the external static 
magnetic field. Since our approach requires two AC field contributions, $\textbf{h}_{\mathrm{AC}}^{\mathrm{cavity}}$ denotes the AC field from the cavity port, and $\textbf{h}_{\mathrm{AC}}^{\mathrm{magnon}}=\delta_0e^{i\phi}\textbf{h}_{\mathrm{AC}}^{\mathrm{cavity}}$ the magnon port, driving magnons only. Thus, the relative phase and amplitude ratio are defined as $\phi=|\phi_{\mathrm{cavity}}-\phi_{\mathrm{magnon}}|$ and $\delta_0=\frac{|\textbf{h}_{x,\mathrm{AC}}^{\mathrm{magnon}}|}{|\textbf{h}_{\mathrm{AC}}^{\mathrm{cavity}}|}$. 
\\\\\
We focus on controlling the Kittel mode's coupling strength, 
which represents a special magnetostatic mode with wave vector $\mathbf{k}=0$ and a dispersion $\omega_{\mathrm{m}}=\gamma {H}_{\mathrm{ext}}
\label{H} $ for a sphere \cite{Kittel_1948,Fletcher_1959}. 
In order to achieve level merging, the modulus of the coupling strength also contains a non-zero imaginary contribution because it becomes complex. 
The imaginary part is associated to controlling the system's dissipation via the coupling \cite{Grigoryan2018,Bernier2018}. Thus, the two-tone driven CMP can be described by the non-Hermitian Hamiltonian $
\mathcal{H}_{\rm{sys}}=\hbar\omega\sous{c} a^{\dag}a+\hbar \omega\sous{m}m^{\dag}m+ \hbar g_{\mathrm{eff}}\left(m^{\dag}a+a^{\dag}m\right)+ \hbar \Omega \left(a^{\dag}m\right),$
\label{TC}
where $\omega_c$ is the cavity resonator photon frequency. 
The last term is the intracavity cavity photon-magnon interaction with macroscopic coupling strength $g_{\mathrm{eff}}=g_0 \sqrt{N}$, where $g_0$ is the single spin coupling strength and $N$ is the total number of contributing spins \cite{9783540285731,safo, Boventer2018}. Since the magnon port acts as an indirect drive to the cavity photons via the coupling of the magnons, the last term models this additional drive by the \akuts{driving frequency} $\Omega =g_{\mathrm{eff}}\delta_0 e^{i\phi}$. 
The reflection scattering parameter $S_{11}(\omega)$ can be derived employing Input-Output theory, which yields \cite{9783540285731}: 
\begin{equation}
S_{11}(\omega)=-1+\frac{2\kappa_{e,1}- \frac{2ig_{\mathrm{eff}}\delta_0 e^{i\phi}(1+\delta_0 e^{i\phi})\sqrt{\kappa_{e,1}\kappa_{e,2}}}{-i(\omega-\omega_m)+\kappa_m}}{-i(\omega-\omega_c)+\kappa_r+\frac{g_{\mathrm{eff}}^2(1+\delta_0 e^{i\phi})}{-i(\omega-\omega_m)+\kappa_m}},
 \label{S11Expression}
\end{equation}
where $\kappa_{e,1},\kappa_{e,2},\kappa_r$ and $\kappa_m$ denote the dissipation parameters due to the coupling of the feedline into the resonator at the magnon and cavity port, the total resonator losses, and the magnon linewidth, respectively. 
Comparing the expression for a one port driven cavity photon-magnon polariton \cite{Boventer2018} with \eqnref{S11Expression}, we define the dependence on $\phi$ and $\delta_0$ of the effective coupling strength as 
 \begin{equation}
{g^\prime (\delta_0,\phi)}={g_{\mathrm{eff}}} \sqrt{1+\delta_0 e^{i \phi}},
 \label{gprime}
\end{equation}
where the value of $g_{\mathrm{eff}}$ corresponds to a complete suppression of the magnon drive ($\delta_0=0$). Accordingly, for certain combinations of $\delta_0$ and $\phi$, the coupling strength contains an imaginary contribution.\\\\ 
 \begin{figure}[h!]
\centering
\includegraphics[scale=0.35]{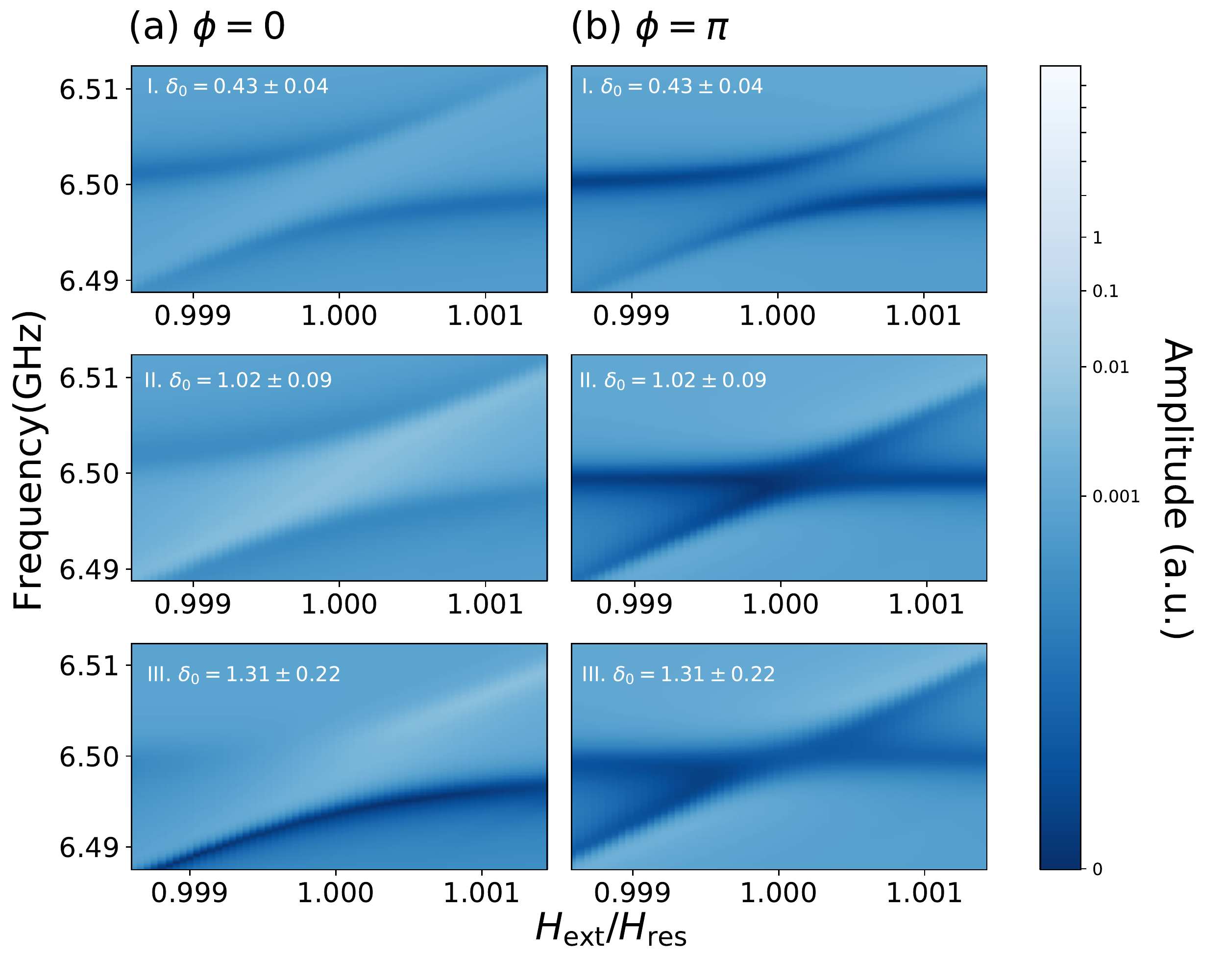}
\caption{Experimental spectra for (a) $\phi=0$ (b) and $\phi=\pi\,$ for three different regimes of $\delta_0$: In I we see level repulsion [$\delta_0<1$], in II the transition [$\delta_0 \simeq 1$], and in III level merging [$\delta_0>1$].
The different combinations of values for $\phi$ and $\delta_0$ lead to different effective couplings, i.e. $g^\prime(\delta_0,\phi)$. Note that level repulsion is always observed for $\phi = 0$, although when $\delta_0>1$ some asymmetry appears. This is attributed to crosstalk in the system. All plots are normalized by the mean value of the signal’s background amplitude and displayed in units of the resonance field $H_{\mathrm{res}}$ which is the point where $\omega_c = \omega_m$ ($H_{\mathrm{ext}}=\textit{\textbf{H}}_{\mathrm{ext}}\hat{\mathbf{z}}$).
}
\label{Spectrum9db}
\end{figure}
In order to study the effect of either changing $\delta_0$ or $\phi$ and keeping the other parameter constant on $g^\prime(\delta_0,\phi)$, we record the dispersion spectra, i.e. frequency dependent response as a function of an externally applied magnetic field for each combination of $\delta_0$ and $\phi$. 
In \figref{Spectrum9db}, we summarize the key features of this work by showing the spectra for relative phase shifts of (a) for $\phi=0$  and (b) for $\phi=\pi$ for three different regimes of $\delta_0$. These are: Below level merging ($\delta_0=0.43\pm 0.07$), part I), at the transition to level merging ($\delta_0=1.02\pm 0.09$, part II), and in the regime of level merging ($\delta_0=1.31\pm 0.22$, part III). 
\begin{figure}[t!]
\centering
\includegraphics[scale=0.4]{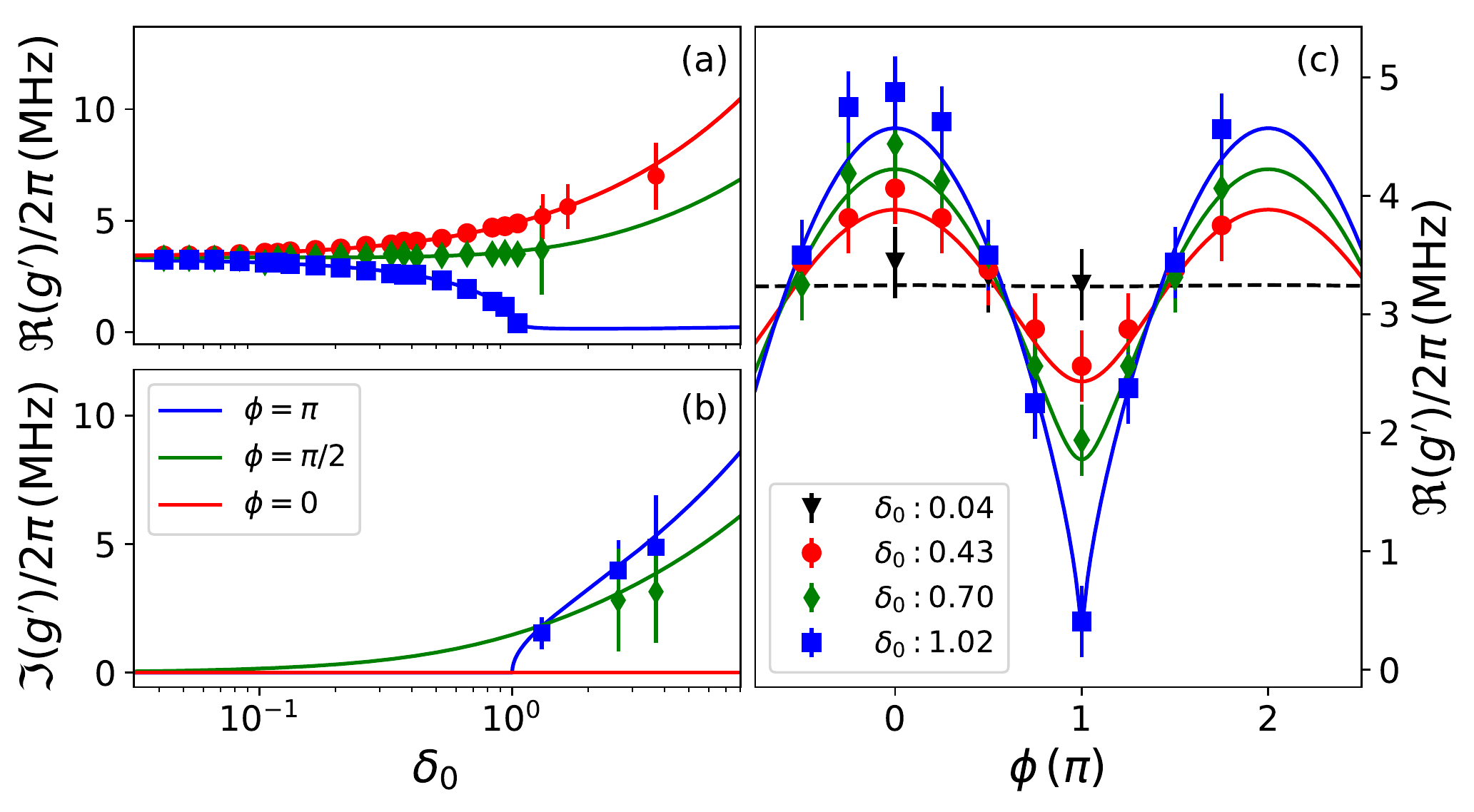}

\caption{Dependence of the real (a) and imaginary part (b) of the coupling strength on $\delta_0$ for
$\phi \in \{0,\pi/2,  \pi$\}. 
For all cases, the real part behaves in a similar manner at very low $\delta_0$ and has the same value within the error bars. For the cases of $\phi = 0$ (red circles) and $\phi = \pi/2$ (green diamonds) the coupling strength increases. However, higher values of the coupling strength are seen for $\phi = 0$. On the other hand, for $\phi=\pi$ (blue squares), we see a decrease towards the limit $\delta_0 \rightarrow 0$ where the coupling strength disappears. This indicates the transition to level merging of the cavity photon and magnon dispersions. This behavior is related to the imaginary part shown in (b). 
For $\phi=\pi/2$ this increase is suppressed as real (level repulsion) and imaginary (level attraction) part are non-zero. 
The solid lines are fits of \eqnref{gprime} to the data. 
(c) Experimental data (points) and fit (solid lines) of $g^\prime (\delta_0,\phi)|_{\delta_0=\mathrm{const}}$ for four different values of $\delta_0$. 
The final value to $\delta_0$ is calculated from the geometric average from both fit results (c.f. \cite{Supplement}). }
\label{powerdep}
\end{figure}
Apart from small crosstalk contributions, the magnon port interacts only indirectly by coupling of the magnons to the cavity photons. 
Here, the amplitude of the AC magnetic field associated to the cavity port is typically higher than that of the magnon port. Therefore, it is necessary to attenuate the cavity port for reaching the regime of level merging.  
As expected from Eq. \eqref{gprime}, for $\phi=0$, the anticrossing gap increases towards higher $\delta_0$.  
\\
\noindent
The AC magnetic field from the magnon port exerts an additional torque on the precessing magnetization where its orientation depends on $\phi$ \cite{Grigoryan2018}. 
Furthermore, the transmission coefficient's amplitude and linewidth depend on the interplay (controlled by $\delta_0$ and $\phi$) between individual dissipation and coupling strength between the two oscillators. 
For the specific case of level merging ($\phi=\pi,\,\delta_0=1$) and on-resonance, this additional torque compensates the intrinsic damping and coupling-induced linewidth broadening. 
Since the coupling strength is a measure for the exchange of energy between cavity photons and magnons, it can be considered as an additional channel for energy dissipation for one subsystem or gain for the other one within one oscillation period for the exchange of energy. As mentioned above, depending on the orientation of the effective acting torque (with contributions by both tones), the damping of the magnons is either enhanced or compensated for  
$\delta_0=1.02\pm 0.09$ [c.f. \figref{Spectrum9db} (b) part II]. 
Hence, the magnon port starts serving as an additional drive for the cavity photons, and a strong enhancement of the amplitude of $S_{11}(\omega)$  and decrease of the linewidth due to level merging is expected at resonance, as we will see later on.  
\newline
In the regime of level merging for $\phi=\pi$, the imaginary part of the coupling term dominates the changes, resulting in a coalescence of the antisymmetric and symmetric solution [c.f. \figref{Spectrum9db} (b) part III].
In \figref{powerdep}, we show the dependence of $g^\prime(\delta_0,\phi)$ as a function of $\delta_0$ and $\phi$ in combination with a fit according to \eqnref{gprime} (solid lines). 
As illustrated in \figref{powerdep} (a)-(b), we display the dependence on $g^\prime(\delta_0,\phi)|_{\phi=\mathrm{const}}$ for $\phi \in \{0,\pi/2,  \pi\}$ which shows the highest difference in terms of the spectrum's shape. 
For $\delta_0 \rightarrow 0$, the influence of the relative phase shift on the coupling strength should decrease until $g^\prime( \delta_0,\phi)={g_{\mathrm{eff}}}$, which is observed as the merging of the three different curves with $\phi $ fixed in \figref{powerdep}. To this end, the magnon port is so strongly attenuated, that the prevailing cavity photon driven cavity photon-magnon coupling dominates again, observed as an anticrossing with a gap of $2{g_{\mathrm{eff}}(\delta_0 \rightarrow 0)}$.   
Now, depending on the specific combination of $\delta_0$ and $\phi$, we transform the coupling strength into a complex quantity where we measure the modulus in the experiment. Since for $\phi=0,\,\Im{g^\prime(\delta_0,\phi)=0}$ for all $\delta_0$, and $\Re{g^\prime(\delta_0,\phi)=0}$ for $\phi=\pi$ and $\delta_0>1$, the real part can be attributed to a repulsive interaction (level repulsion) and the imaginary part to an attractive one (level merging), respectively. Thus, in accordance with the expectation from \eqnref{gprime} [c.f. \figref{powerdep} (a)], the coupling strength for $\phi=0$ is maximized towards $\delta_0=1\,(1.02\pm 0.09)$ with a total increase of $\approx 33\%$. 
As it is shown in \figref{powerdep} (a), for $\phi=\pi$ and $\delta_0<1$, the contribution from the additional torque acting on the magnetization increases and results in continuously decreasing the value of the anticrossing gap, i.e. real value of the coupling strength.   
Beyond the transition to level merging, [$\phi=\pi$ and $\delta_0>1$], we can also extract the imaginary contribution \cite{Supplement}. The width of the region for level merging corresponds to $4\Im{g^\prime(\delta_0,\phi)}$ and $\Re{g^\prime(\delta_0,\phi)}=0$ [c.f. \figref{powerdep} (b)] \cite{Bernier2018}. However, 
for $\phi=\frac{\pi}{2}$, we can also observe a co-existence of anticrossing and level merging, resulting in a lower increase of $\Re{g^\prime(\delta_0,\phi=\pi/2)}$. Hence, it demonstrates the broad tunability of our system. 
\figref{powerdep} (c) shows the dependence of the coupling strength on $\phi$ for $\delta_0$. 
Within the error bars, the points for $\delta=0.04$ are independent of the relative phase. 
It confirms the previous observation (\figref{powerdep}), that for $\delta_0 \rightarrow 0$, the magnon port's contribution is strongly suppressed regardless of the current value for $\phi$. There, the system with two drives enters a regime for $\delta_0$, where it can be effectively described as a single-tone driven CMP with coupling strength $g_{\mathrm{eff}}$.
For a noticeable contribution from the magnon port but yet not large enough to reach the regime of level merging ($\delta_0 < 1$), the gap is decreased without vanishing completely [c.f. \figref{powerdep} (c) $\delta_0=0.43\pm 0.07$]. The complete merging of the anticrossing gap formed from the hybridized cavity photon magnon system is observed 
at  $\delta_0=1$ [see blue squares in Fig. 3 for $\delta_0=1.02\pm 0.09$]. In this regard, the coupling strength changes by nearly an order of magnitude from   $g_{\mathrm{eff}}(\delta_0,\phi)/2\pi|_{\delta_0 =0.04} =(3.43\pm 0.3\,\mathrm{MHz})$ to the smallest observed value of  $g^\prime(\delta_0,\phi)/2\pi=(0.37\pm 0.3 \,\mathrm{MHz})$. For $\phi=0$, the increase in coupling strength is less than the total decrease at $\phi=\pi$, because here the coupling strength is \akuts{just} increased by $g^\prime(\delta_0,0)\propto\sqrt{(1+\delta_0)}$.
\begin{figure}[t!]
\centering
\includegraphics[scale=0.53]{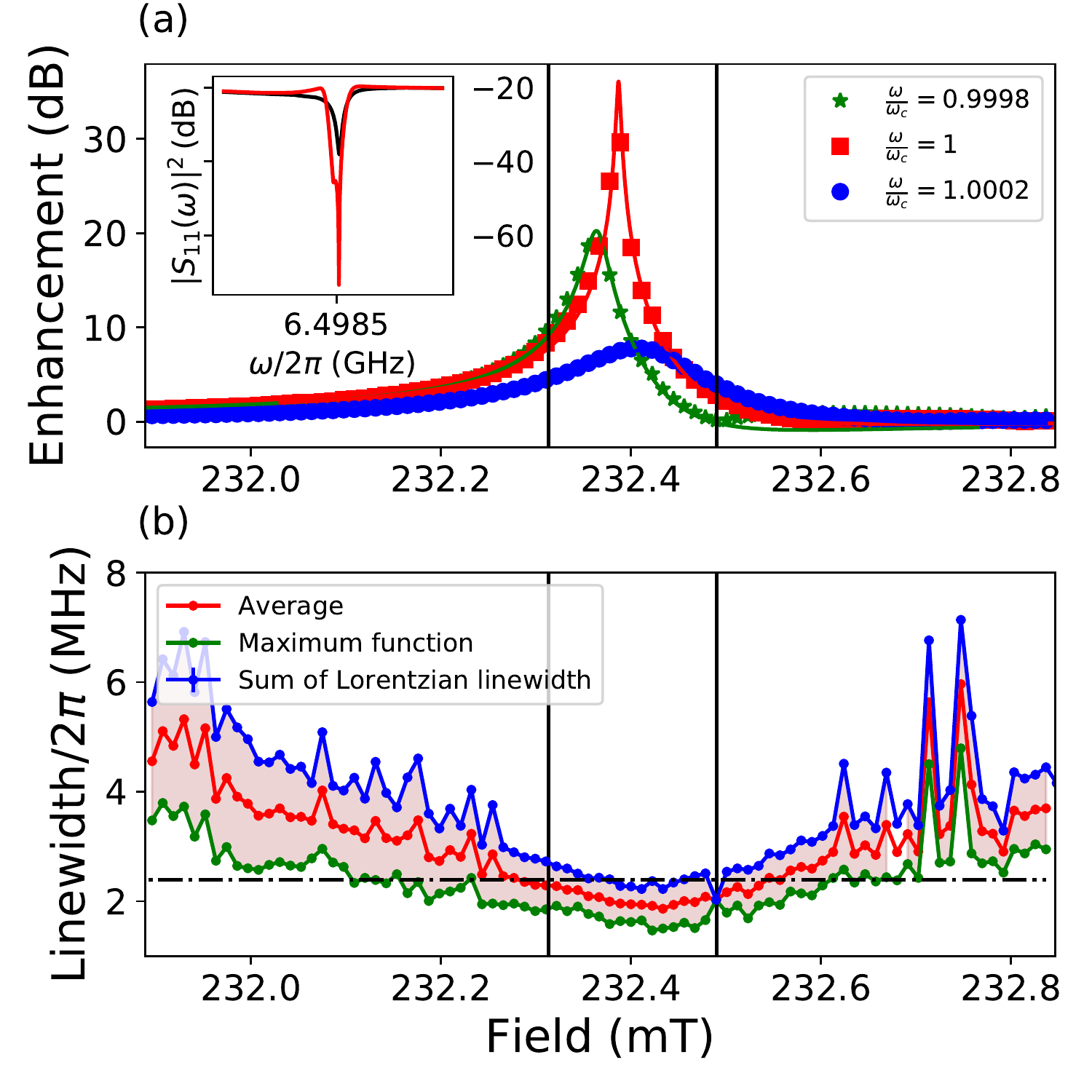}
\caption{(a) Amplitude  enhancement at level merging ($\delta_0=1.02\pm 0.09, \phi=\pi$) corresponding to the absolute difference between the peak amplitudes of off ($H_{\mathrm{ext}}\neq H_{\mathrm{res}}$) and on ($H_{\mathrm{ext}}=  H_{\mathrm{res}}$) resonant signals at the cavity resonance frequency ($\omega_c/2\pi=6.4985\,\mathrm{GHz}$). The inset shows the off (black)- and on (red) response for ($\mu_0H_{\mathrm{res}}\approx 232.4\,\mathrm{mT}$) showing the maximal enhancement of the amplitude. For a field interval of $\approx \pm 0.1\,\mathrm{mT}$ around resonance, the amplitude is strongly enhanced, with a maximal value of $30\,\mathrm{dB}$ and the linewidth further decreased due to level merging (vertical black lines). The small kink to the left of the on-resonant signal is due to the asymmetry in the level merging spectrum [c.f. \figref{Spectrum9db}(b) II]. The solid lines correspond to a fit to \eqnref{S11Expression} at fixed frequency. 
(b) Field dependence of the average linewidth value (red), estimated  between a lower bound (green) by the maximum function and an upper bound (blue) by a sum of the linewidth of two Lorentzian functions yielding a certain interval (shaded). The dashed line (black) refers to the geometric mean of the individual cavity photon and magnon linewidth. For a linewidth decrease due to level merging, the minimal linewidth around resonance should be lower. Due to level merging, the linewidth falls below that geometric mean and further decreases by $\approx 13\%$.  }  
\label{amplitude_enhancement}
\end{figure}\newline
For an CMP created by the cavity port only, an increase of the signal's linewidth at resonance has been reported \cite{Harder2018a}. However, for our specific two-tone system, a decrease in linewidth, accompanied by a strong enhancement of the resonance amplitude is expected to emerge in the regime of level merging \cite{Grigoryan2018}. 
Indeed, as shown in \figref{amplitude_enhancement}, we observe an enhancement of the amplitude and decrease in linewidth at the transition of level merging $\delta_0=1$ and $\phi=\pi$. 
The amplitude at resonance of the coalesced anticrossing at $\phi=\pi$ is increased by $30\,\mathrm{dB}$. 
In \figref{amplitude_enhancement} (a), we show the relative increase of the amplitude compared to the off-resonant cavity resonator's amplitude ($H_{\mathrm{ext}} \ll H_{\mathrm{res}}$) below, at, and above the cavity's resonance frequency as illustrated in the inset for the off- and on-resonance signal. Since the CMP can be regarded as the quasiparticle from a system of two coupled harmonic oscillators, the linewidth is found by fitting the sum of two Lorentzian functions to the data and determined by the geometric mean between a lower and upper bound. While the lower bound is given by the Maximal function, which always takes the higher value of the set of both linewidths \cite{EliasM.Stein2005}, the upper bound is given by the sum of the individual linewidths. Thus, for a decrease of the linewidth due to level merging, this average value of the linewidth needs to be below the average of the off-resonant linewidths of magnon and cavity photon. They are determined to $\kappa_r/2\pi=3.79\pm 0.003\,\mathrm{MHz}$ and $\kappa_m/2\pi=1\pm 0.5\,\mathrm{MHz}$, i.e. the linewidth has to be below its geometric mean of $2.4\pm 0.25\,\mathrm{MHz}$. We observe an additional decrease of the linewidth by $\approx 13\,\%$ since the mean of the values below this value is $2.12\pm 0.21\, \mathrm{MHz}$ around resonance (black, vertical lines in the figure). This change is attributed to the decrease of the linewidth due to level merging. 
\newline
In summary, we demonstrated a method which allows us achieving a full control of the coupling strength in hybrid cavity photon-magnon polariton systems. This is done by simply tuning the relative phase $ \phi$ and amplitude $\delta_0$ between the cavity photons and magnon external drive (added through a second port). By controlling such parameters we observe a full collapse of the gap of the anticrossing at resonance, a regime we call \akuts{level merging}. This is observed only if the relative phase is set to $\phi =\pi$ as well as the relative amplitude ratio to $\delta_0=1$ \cite{Doerflinger2018}. We note that this transition, mediated by the two-toned system, is particularly interesting as it can be used to enhance the amplitude of the signal. 
 Moreover, our system realizes a fully automated tuning mechanism wherein we can easily shift through various levels of coupling, including the recently studied level attraction \cite{Harder2018a} (using higher $\delta_0$’s). In our system, however, this is done without any direct changes of the experimental setup, such as moving the sample, which reduces error. Such automated control mechanism over the spin-photon interaction could pave the way for deliberately turning on and off the coherent exchange of information. Accordingly, this could enable future applications for data storage and information processing by the addition of a non-linear component such as a superconducting circuit to the spin-photon system.
\\\newline
We acknowledge valuable discussions with Can-Ming Hu, Bimu Yao, Vahram L. Grigoryan, Ka Shen, and Ke Xia. This work is supported by the European Research Council (ERC) under the Grant Agreement 648011 and the DFG through SFB TRR 173/Spin+X. R. Mac{\^e}do acknowledges the support of the Leverhulme Trust. T. Wolz acknowledges financial support by the Helmholtz International Research School for Teratronics.
\bibliography{LMP_Final_Ref}
\bibliographystyle{apsrev}
\newpage

\section*{Supplementary Material}
\noindent
\subsection{Determination of $\delta_0$}
\label{SupI}
\noindent
In the current setup, it is not possible to directly measure the amplitudes of the internal alternating current (AC) magnetic fields from the magnon and the cavity port, respectively. 
Thus, experimentally, the internal amplitude ratio of $\delta_0$ cannot be directly determined. 
Instead, we can only determine an external relative amplitude ratio $\delta_{\mathrm{ext}}$ before the signal is coupled to the cavity port and magnon port, respectively. 
We relate it to the internal amplitude ratio via $\delta_0=\zeta\delta_{\mathrm{ext}}$, where, $\zeta $ denotes an effective coupling factor for the coupling into the cavity, determined from fitting for $g^\prime(\delta_0,\phi)|_{\phi=\mathrm{const}}$ (c.f. \figref{powerdep}(a)). From fitting to $g^\prime(\delta_0,\phi)|_{\delta_0=\mathrm{const}}$ (c.f. \figref{powerdep} (c)), values for $\delta_0$ are also found which are compared to the previous fit result.
As can be inferred from \tabref{delta_comp}, the values for $\delta_0$ are very close. 
\begin{table}
\centering
\begin{tabular}{|cc|c|c|c|c|c|c|}
\hline 
(a)&$\delta_0$ & 0.436 & 0.555 & 0.700 & 0.867 & 0.913 & 0.99 \\ 
\hline 
(b)&$\delta_0$ & 0.417 & 0.525 & 0.661 & 0.832 & 0.923 & 1.047 \\ 
\hline 
(c)&$\overline{\delta_0}$ & 0.426 & 0.540 & 0.681 & 0.850 & 0.923 & 1.019 \\ 
\hline 
(d)&$\delta_{\mathrm{ext}}$ & 1 & 1.259 & 1.585 & 1.995 & 2.239 & 2.512 \\ 
\hline 
\end{tabular} 
\caption{Comparison of the (fit) results for the amplitude ratio. (a) Values for $\delta_0 $ determined from a direct fit to $g^\prime(\delta_0,\phi)|_{\delta_0=\mathrm{const}}$. (b) Values for $\delta_0$ calculated from the values for $\delta_{\mathrm{ext}}$ and the determination of $\zeta$ from a fit to the data from $g^\prime(\delta_0,\phi)|_{\phi=\mathrm{const}}$. (c) Geometric mean for the results from (a) and (b). (d) Calculated values for the external amplitude ratio $\delta_{\mathrm{ext}}$. }
\label{delta_comp}
\end{table} 
The value for $\zeta$ is determined to $\zeta=0.418\pm 0.065$. In order to minimize the error on the determination of $\delta_0$ the geometric average including the error is calculated. The result is also given in \tabref{delta_comp} and these values for $\delta_0$ are utilized in the discussion of the results and distinction of the regimes for anticrossings or occurrence of level merging. 
Note, that in both cases, the errors are calculated from the covariance matrix. Then, the geometric mean for the result for $\zeta$ for $\phi=0,\pi/2,\pi$ and the error of the mean value utilizing Gaussian error propagation are calculated.
This procedure results in $\zeta=0.418\pm 0.065\approx 0.42\pm 0.07$. 
For the data with higher $\delta_0$, the respective value and corresponding error again employing Gaussian error propagation are calculated from the relation $\delta_0=\zeta\delta_{\mathrm{ext}}$. 

\subsection{Procedure for data analysis}
\label{SupII}
\noindent
In order to study the dependence of the coupling strength on $\delta_0$ and $\phi$, the center frequencies of the symmetric (lower, $\omega<\omega_0$) and antisymmetric (upper, $\omega>\omega_0$) branch at resonance need to be precisely determined, where $\omega_0$ denotes the resonance frequency of the resonator, i.e., the uncoupled system. Since the anticrossing gap $\Delta \omega$ corresponds to $2g^{\prime}(\delta, \phi)$, the distance between upper and lower branch is determined and yields the coupling strength. However, in this specific experiment, two features are present, which makes it necessary to include the phase of the respective signal as well.\\
First, the dispersion spectra are recorded in an $S_{11}(\omega)$ reflection measurement at the cavity port. 
Since a reflection measurement is recording losses of a microwave input signal at the device under test (DUT), the variation of $\delta_0 $ does not only change the general baseline value but as well the signal to noise ratio (SNR) of the cavity signal to the background. 
Considering the amplitudes at resonance, we find either vanishing peaks into the background or ambiguity in separating peaks from the background signal. Second, the small contribution from crosstalk results in measuring a small transmission signal from the magnon to the cavity port superimposed to the reflection signal of interest. Then, especially as the cavity resonator is further attenuated which again worsens the SNR, the total measured amplitude can become really small or even change sign for a strongly attenuated cavity resonator as it can be seen for instance in \figref{Spectrum9db} b.II). Thus, analyzing the amplitude data only is not sufficient, and the phase data is also considered. First, the phase delay and slope are corrected for distortions due to contributions such as cables of a certain length. In our data, the position of the peak in the amplitude frame corresponds to the point of maximal slope in the phase spectrum. Consequently, the gradient of the phase is calculated and an algorithm for finding the corresponding peaks is applied to the data for different values of the externally applied field. Once the peak positions are known, the frequency difference between them is calculated.
\subsection{Estimation of crosstalk}
\noindent
\begin{figure}
\centering
 \includegraphics[scale=0.2]{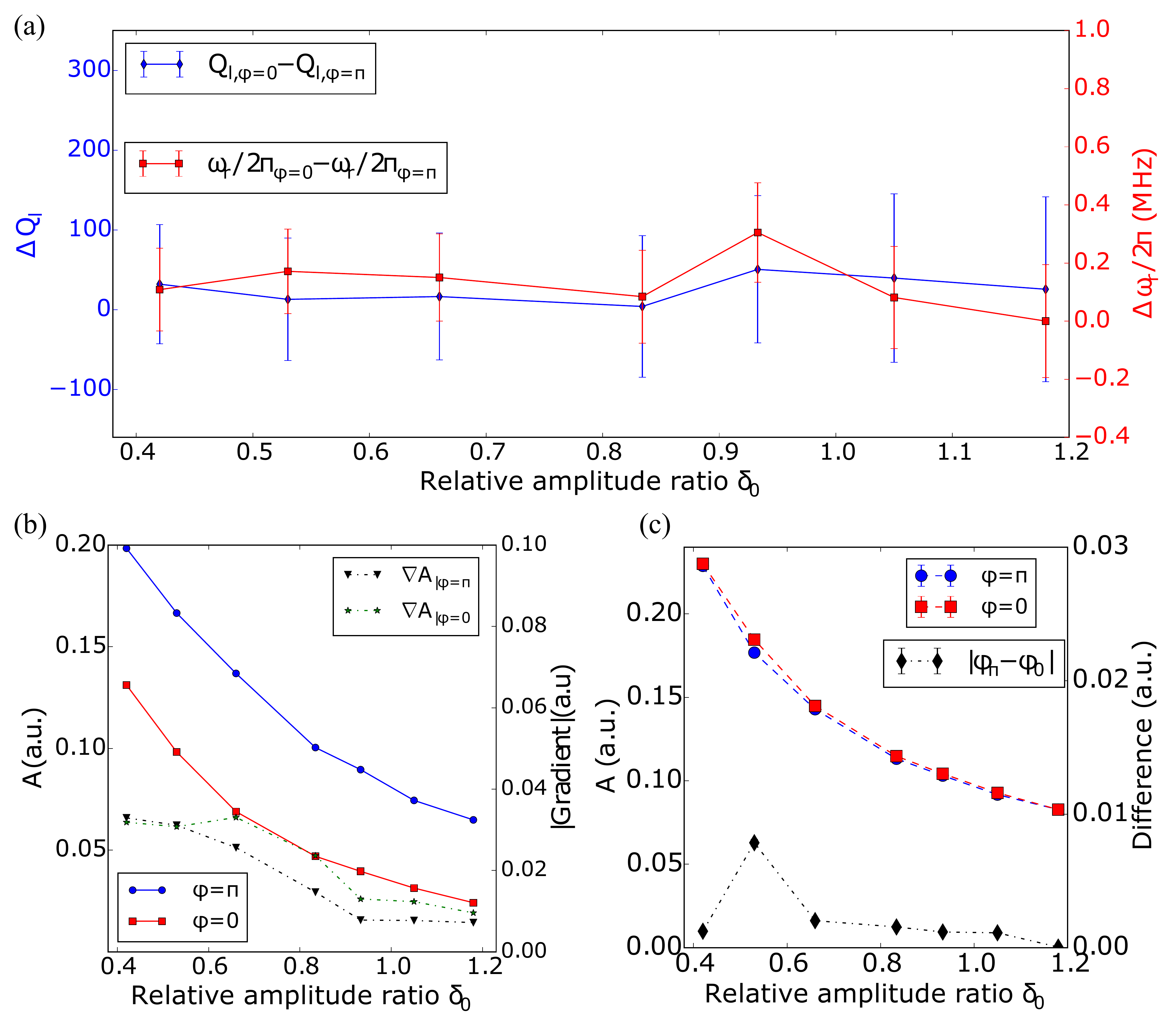}
 \caption{
 Examination of the influence of crosstalk due to a phase shift of $\Delta \phi=\pi$ on the off-resonant 
($ H_{\mathrm{ext}} \ll H_{\mathrm{res}}$) cavity resonator signal with respect to loaded quality factor, resonance frequency ((a)), cavity signal [$\omega=\omega_c$, (b)] and baseline amplitude [$\omega \neq \omega_c$, (c)].  
(a) Dependence of the difference of the loaded quality factor $Q_l$ for $\phi=0$ and $\phi=\pi$ on $\delta_0$. Within the error bars, the difference to $Q_l$ and $\omega_c$ remains constant as a function of changing the relative amplitude ratio $ \delta_0$. On average, the crosstalk leads to a change of $\Delta\overline{Q_l}=25$ and resonance frequency shift of $\Delta\overline{\omega_c}/2\pi=0.18 \,\mathrm{MHz} \,(0.006 \,\mathrm{mT})$.  (b) Off-resonant change of the amplitude A due to a sweep of the relative phase from 0 to $\pi$. The values displayed are the amplitudes, corrected by the mean of the background signal. The general decrease in amplitude is due to the increased value of the attenuator. Up to a small decrease, the crosstalk remains almost constant as the gradient for $\phi=0$ and $\phi=\pi$ does not change significantly. 
The more the cavity port is attenuated,  the higher is the relative contribution from the second, the magnon port in form of transmission measured at the cavity port. Hence, the crosstalk's influence increases and can distort the spectra for high values of $\delta_0$. (c) Average over the baseline values at the recorded frequency window for two different phases. A phase shift of $\pi$ does not change the baseline, the data points cover the error bars for the difference. }
 \label{Crosstalk}
\end{figure}   
The crosstalk is the direct excitation of cavity photons by the microwave energy at the magnon port. It reduces the contribution to the relative amplitude $\delta_0$ from the magnon port. On the other hand, the recorded reflection amplitudes at the cavity port and the resonance frequency can be strongly altered for a sweep of the relative phase $\phi$ from 0 to $\pi$. 
This impedes a straightforward analysis of the phase-dependent changes of the coupling strength and variation of the relative amplitude of  $\delta_0$. 
By means of comparing the off-resonant cavity resonance dip for the extremal cases of $\phi=0$ and $\phi=\pi$, an estimate for the crosstalk contribution can be made. 
We evaluate it by examining the change in the loaded quality factor $Q_l$, which is a measure for both internal and external cavity resonator losses and in the resonance frequency. Besides, we compare the difference in total peak height over the background amplitude. Correspondingly, \figref{Crosstalk} displays the difference in quality factor $Q_l$ and resonance frequency $\omega_c$ [(a)] and the variation from off- to on- resonant amplitudes between the two-phase settings $\phi=0$ and $\phi=\pi$ [(b)]. The error bars in \figref{Crosstalk} (a) are a result of a circle fit \cite{probst, QKIT} to the resonator and Gaussian error propagation due to the calculation of the difference. Within these error bars, both loaded quality factor $Q_l$ and the cavity resonator's resonance frequency $\omega_c$, remain constant. Averaged over the different attenuation values yields differences of  $\Delta\overline{Q_l}=25$ and $\Delta\overline{\omega_c}/2\pi=0.18\,\mathrm{MHz}\, (0.006 \,\mathrm{mT})$ for the loaded quality factor $Q_l$ and the resonance frequency of the off-resonant cavity response, respectively. Therefore, the crosstalk itself is independent of sweeping $\delta_0$. Further, the change in the cavity resonator's losses and especially the shift in resonance frequency due to crosstalk are negligibly small and do not alter the level merging at $\delta_0=1.02\pm0.09,\, \phi=\pi$ dramatically. 
The difference in total height of the peak increases towards a higher attenuation on the cavity path since relative to the contribution from the cavity port, the crosstalk contribution from the second, the magnon port's
input is enhanced. The enhancement exactly corresponds to the attenuation of the cavity path of $9\,\mathrm{dB}$ ($\delta_0=1.18\pm 0.19$). 
\subsection{Determination of the imaginary part}
\label{SupIV}
\begin{figure}
\centering
\includegraphics[scale=0.5]{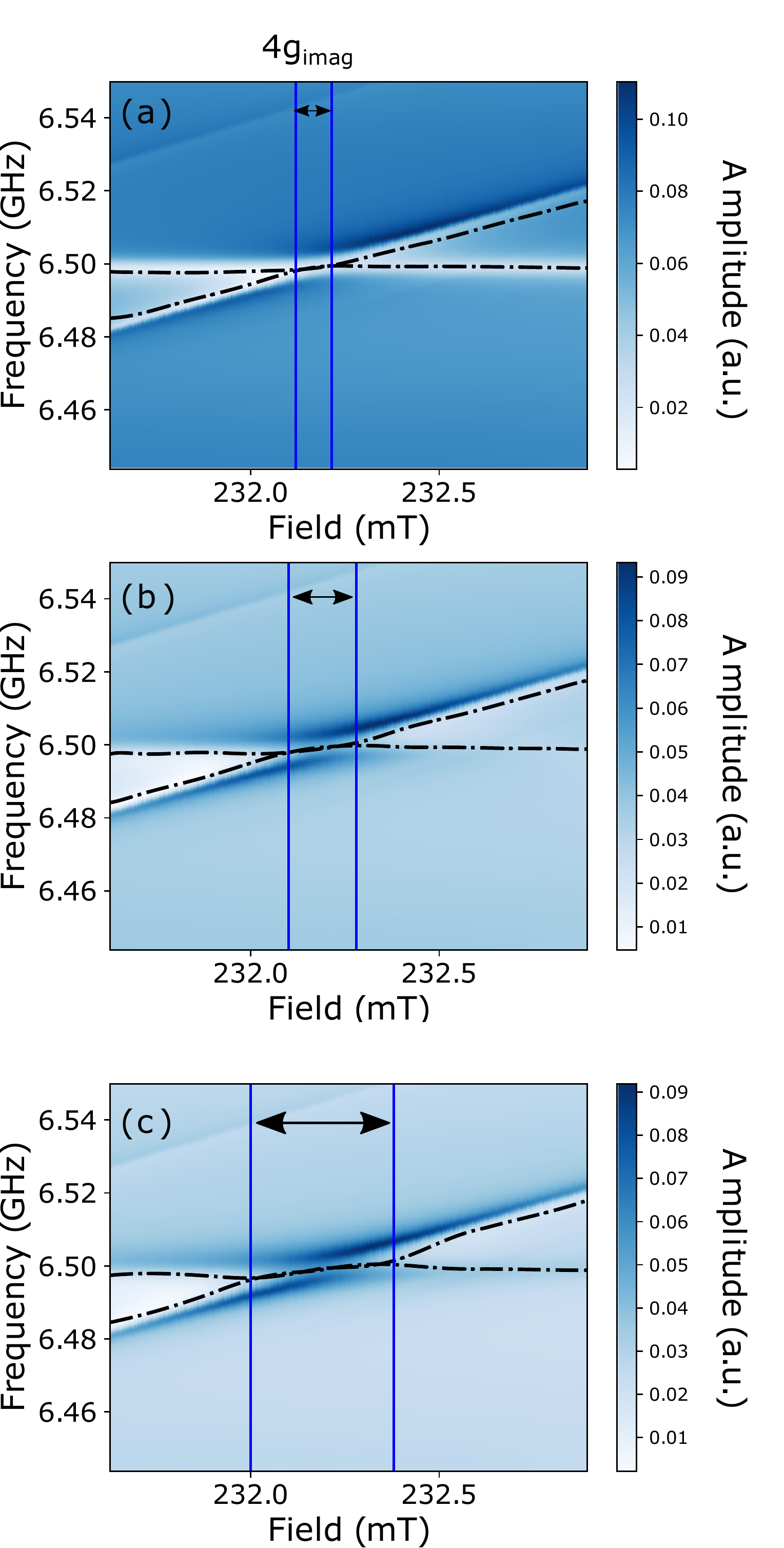}
\caption{Spectra plus identified peaks (black, dotted line) for $\delta_0=1.31$ [(a)], $\delta_0=2.15$ [(b)], and $\delta_0=2.71$ [(c)]. We found from simulations, confirming a finding from Ref. \cite{Bernier2018}, that this distance corresponds to $4\Im{g^\prime(\delta_0,\phi)}$. As the cavity port is more and more attenuated, the relative strength of the contribution from direct crosstalk increases which is seen by the anticrossing like behavior around resonance.  }
\label{GIM}
\end{figure}

The values for the imaginary part in \figref{powerdep} (b), are determined by fitting a sum of two Lorentzian functions to our data. As the peaks are approaching each other, we set a threshold until the peak distance is not resolvable any more. The vertical distance between the apexes of both triangles is found by identifying the width in terms of the applied external magnetic field where the peak distance is below the threshold. 
Since this value strongly depends on the chosen threshold, the given values correspond to the mean value from three different thresholds. One corresponds to the lower limit of the total linewidths which is the Kittel mode's linewidth, the second for an intermediate value and the third the sum of both linewidths as an upper bound because above the peaks can are distinguishable. The result is shown in \figref{GIM}. 
\subsection{Derivation of scattering formula}
\label{SupV}
We start the derivation from the system Hamiltonian describing the intracavity components and interactions between the constituents. 
In addition, similar to a modification in Ref. \cite{Fischer2018} and Ref. \cite{Zou2013}, we modify the standard Tavis-Cummings Hamiltonian to a driven one. The drive originates from the indirect coupling of additional energy by the second port, which is only coupled to the magnons, which are excited by an effective AC magnetic field, comprised from the individual contributions of each port. 
By the coupling of the magnons to the cavity photons, the cavity photons experience another internal drive by coupling. The second port is incorporated here.
\begin{eqnarray}
\mathcal{H}_{\mathrm{sys}}=\hbar\omega\sous{c} a^{\dag}a+\hbar \omega\sous{m}m^{\dag}m+ \hbar g_{\mathrm{eff}}\left(m^{\dag}a+a^{\dag}m\right)+\hbar \Omega a^{\dag}m,
\end{eqnarray}
where $\Omega=g_{\mathrm{eff}}\delta_0 e^{i\phi}$. 
The last term denotes the additional input from the magnon port via the coupling of the magnons to the cavity photons. The driving $\Omega$ frequency depends on the effective coupling strength and the relative phase and amplitude of the second tone of the magnon port, which it only drives. A direct excitation, i.e., cavity photons from the magnon port microwave photons refers to crosstalk which is suppressed as much as possible in the experiment. Instead, the magnon port serves only as a direct source of exciting the magnetization precession, i.e., the magnon. It is only indirectly seen by the cavity by the coupling strength, opening another channel depending on the relative phase and amplitude of energy exchange. 
The equations of motion are written down in Langevin form, resulting in:
\begin{equation}
\frac{\partial m(t)}{\partial t}=-i\omega_m m(t)-ig_{\mathrm{eff}}a(t)-\kappa_m m(t)+\sqrt{2\kappa_{e,2}}b_{\mathrm{in},2}(t)
\end{equation} and
\begin{equation}
\frac{\partial a(t)}{\partial t}=-i\omega_c a(t)-ig_{\mathrm{eff}}(1+\delta_0 e^{i\phi})m(t)-\kappa_r a(t)+\sqrt{2\kappa_{e,1}}b_{in,1}(t),
\end{equation}
where $b_{\mathrm{in},2}(t)=\delta_0 e^{i\phi} b_{\mathrm{in},1}(t)$.
After a Fourier transform, expressing the equations in terms of a frequency dependence and using of the input-output relation for a system with one external port $b_{\mathrm{out}}+b_{\mathrm{in},1}=\sqrt{2\kappa_{e,1}}a$ ,  we write for the resulting scattering parameter \cite{9783540285731}:
 \begin{eqnarray}
S_{11}(\omega)=-1+\frac{2\kappa_{e,1}}{Y+\frac{g_{\mathrm{eff}}^2(1+\delta_0 e^{i\phi})}{X}}-\frac{2ig_{\mathrm{eff}}\delta_0 e^{i\phi}(1+\delta_0e^{i\phi})}{X \left(Y+\kappa_r+\frac{g_{\mathrm{eff}}^2(1+\delta_0 e^{i\phi})}{X}\right)},
 \end{eqnarray}
 where $X=-i(\omega-\omega_m)+\kappa_m$ and $Y=-i(\omega-\omega_c)+\kappa_r$.
Note, that we measure in reflection at the cavity port, where we also excite the cavity photons of our specific cavity mode and, here,  neglect crosstalk for the moment,
\end{document}